\documentclass[a4paper,12pt]{article}
\usepackage{amsmath}
\usepackage{amssymb}
\usepackage{latexsym}
\usepackage{enumerate}
\newcommand{\be}{\begin{eqnarray}}
\newcommand{\ee}{\end{eqnarray}}

\newcommand{\bsub}{\begin{subequations}}
\newcommand{\esub}{\end{subequations}}

\newcommand{\disfrac}[1][2]{\displaystyle\frac}
\begin{document}

\title{Distribution of Energy-Momentum in a Schwarzschild-Quintessence Space-time Geometry}
\author{Irina Radinschi$^{\text{*1}}$, Theophanes Grammenos$^{\text{**2}}$ and Andromahi
Spanou$^{\text{***3}}$ \\
$^{\text{1}}$Department of Physics \\
``Gh. Asachi'' Technical University, \\
Iasi, 700050, Romania\\
$^{\text{2}}$Department of Civil Engineering, \\
University of Thessaly, 383 34 Volos, Greece\\
$^{\text{3}}$School of Applied Mathematics and Physical Sciences,\\
National Technical University of Athens, 157 80, Athens, Greece\\
$^{\text{*}}$radinschi@yahoo.com, $^{\text{**}}$thgramme@uth.gr,\\
$^{\text{***}}$aspanou@central.ntua.gr}
\date{}
\maketitle

\begin{abstract}
An analysis of the energy-momentum localization for a four-dimensional\break Schwarzschild black hole surrounded by quintessence is presented in order to provide expressions for the distributions of energy and momentum. The calculations are performed by using the Landau-Lifshitz and Weinberg energy-momentum complexes. It is shown that all the momenta vanish, while the expression for the energy depends on the mass $M$ of the black hole, the state parameter $w_{q}$ and the normalization factor $c$. The special case of $w_{q}=-\frac{2}{3}$ is also studied, and two limiting cases are examined.
\end{abstract}

\section{Introduction}

In the field of General Relativity the lack of a unique expression for the
energy density is an important open problem. An efficient definition for the energy-momentum localization has to provide a generally accepted expression for the energy density. Until now, none of the
definitions developed for energy-momentum has given a powerful confirmation that could have been considered the best tool for the energy-momentum localization.

However, there are many proposed tools to calculate the energy-momentum in General
Relativity, such as superenergy tensors \cite{Bel}, quasi-local expressions \cite{Brown} and
the energy-momentum complexes of Einstein \cite{Einstein}, Landau-Lifshitz \cite{Landau},
Papapetrou \cite{Papapetrou}, Bergmann-Thomson \cite{Bergmann}, M\o ller \cite{Moller_1}, Weinberg \cite{Weinberg}, and Qadir-Sharif \cite{Qadir}. These definitions have been applied to many gravitational
backgrounds and have yielded significant results, but almost all of them are coordinate
dependent. In order to avoid the problem of coordinate dependence, the
tele-parallel theory of gravitation \cite{Moller_2} has been used for the calculation
of the energy-momentum distribution in recent years.

Regarding the energy-momentum complexes \cite{Einstein}-\cite{Qadir}, only the M\o ller
energy-momentum complex can be used to evaluate the energy-momentum for a
given space-time in any coordinate system. The other prescriptions are applied
in Cartesian coordinates. The efficiency of the energy-momentum complexes is
sustained by a large number of works. In fact, the pseudotensorial definitions have produced 
reasonable results for various geometries (see, e.g., \cite{Virbhadra} and references therein with an emphasis on Landau-Lifshitz and Weinberg prescriptions), and it is important to point out the similarity of some of these results with those yielded by their tele-parallel versions (see, e.g., \cite{Gamal}). Indeed, Chang, Nester and Chen \cite{Chang} have attempted and succeeded to rehabilitate the energy-momentum complexes.
Different quasi-local definitions correspond to different boundary
conditions.

The remainder of this article is structured as follows: in Section 2 a four-dimensional Schwarzschild black hole surrounded by quintessence is presented. In Section 3 we introduce the Landau-Lifshitz and Weinberg energy-momentum complexes used to calculate the distribution of energy-momentum. Section 4 contains the calculations for the geometry chosen as well as for the specific value $w_{q}=-\frac{2}{3}$ of the state parameter. Finally, in Section 5 we summarize our main results and briefly discuss some particular and limiting cases. Throughout our paper we use for the
calculations the signature ($1,-1,-1,-1$) and geometrized units ($c=1;G=1 $). Greek (Latin) indices take values from $0$ to $3$ ($1$ to $3$).

\section{Schwarzschild Black Hole Surrounded by\break Quintessence}

In the study of the accelerated expansion of the universe, the physical nature of dark energy is one of the most significant and challenging open issues as dark energy is considered the main component of the universe contributing to negative pressure and driving the observed acceleration (besides the alternative of turning, for example, to modified EinsteinÕs field equations). Alternatively to the most conventional dark energy candidate, i.e. the cosmological constant $\Lambda$ (with a state parameter $w = -1$), one important family of other dark energy candidates are scalar field models including, amongst many, quintessence, k-essence (kinetic quintessence, having non-canonical kinetic energy terms), quintom (the equation of state crosses the $w = -1$ boundary from above to below or vice versa), hessence (a new form of quintom), chameleons (self-interacting scalar fields strongly coupled to matter), tachyons, phantoms, dilatons, ghost condensates, and the, different from scalar field models, Chaplygin and generalized Chaplygin gas (for a general review see, e.g., \cite{Amendola}).

In the present work we focus on quintessence, a scalar field minimally coupled to gravity and associated with a potential  $V(\phi)$ decreasing as the scalar field  $\phi$ increases, as the most plausible dynamical vacuum energy model to be combined with the geometry exterior to a black hole. In the last decade, there has been some work in this area (see, e.g., \cite{Kiselev}, \cite{Fernando}, \cite{Zhang}). The question of the energy-momentum localization, by using the M\o ller and Einstein energy-momentum complexes, for a Reissner-Nordstr\"om black hole surrounded by quintessence has already been studied \cite{Saleh}. In order to apply the Landau-Lifshitz and Weinberg energy-momentum complexes, we have the following metric describing the space-time geometry exterior to a Schwarzschild black hole surrounded by quintessence \cite{Kiselev}--\cite{Fernando}:

\begin{equation}\label{line_element}
ds^{2}=
(1-\frac{2M}{r}-\frac{c}{r^{3w_{q}+1}})dt^{2}-
\frac{dr^{2}}{(1-\frac{2M}{r}-\frac{c}{r^{3w_{q}+1}})}dr^{2}-r^{2}(d\theta ^{2}+sin^{2}\theta d\phi
^{2}),
\end{equation}
where $M$ represents the mass of the black hole, $w_{q}$ is the state
parameter of quintessence and $c$ is a normalization factor. The state parameter values lie in the interval 
$-1<$ $w_{q}<-\frac{1}{3}$. For quintessence, the equation of state is $p_{q}=w_{q}\rho _{q}$ with
$\rho _{q}=-\frac{c}{2}\frac{3w_{q}}{r^{3(1+w_{q})}}$, while $c>0$. Thus,  the pressure  $p_{q}$ of
quintessence is negative and the matter energy density $\rho _{q}$ is positive.

For $w_{q}=-1$ the geometry reduces to the Schwarzschild-de Sitter
gravitational background, while for the special choice $w_{q}=-\frac{2}{3}$, the line element (\ref{line_element}) becomes:
\begin{equation}\label{special_line_element}
ds^{2}=(1-\frac{2M}{r}-cr)dt^{2}-\frac{1}{(1-\frac{2M}{r}-cr)}
dr^{2}-r^{2}(d\theta ^{2}+sin^{2}\theta d\phi ^{2}).
\end{equation}
In the case $M<1/8c$, the geometry described by (\ref{special_line_element}) has an inner and an outer horizon \cite{Fernando} given by:
\begin{subequations}
\begin{equation}\label{inner}
r_{\text{inner}}=\frac{(1-\sqrt{1-8Mc})}{2c},
\end{equation}
\begin{equation}\label{outer}
r_{\text{outer}}=\frac{(1+\sqrt{1-8Mc})}{2c},
\end{equation}
\end{subequations}
while in the case $M>1/8c$ the geometry exhibits no horizons and in the case $M=1/8c$, $r_{\text{inner}}=r_{\text{outer}}=1/2c$. The inner horizon corresponds to the Schwarzschild black hole horizon, while the outer horizon corresponds to a cosmological horizon as the one obtained in the case of the Schwarzschild-de Sitter space-time geometry.

\section{The Landau-Lifshitz and Weinberg Energy-Momentum Complexes}

The Landau-Lifshitz energy-momentum complex \cite{Landau} is defined as
\begin{equation}\label{LL}
L^{\mu \nu }=\frac{1}{16\pi }S_{,\,\rho \sigma }^{\mu\nu \rho\sigma},
\end{equation}
where the Landau-Lifshitz superpotentials are given as
\begin{equation}\label{LL_super}
S^{\mu \nu \rho \sigma }=-g(g^{\mu \nu }g^{\rho \sigma }-g^{\mu \rho }g^{\nu
\sigma }).
\end{equation}
The $L^{00}$ and $L^{0i}$ components correspond to the energy and momentum density, respectively. The Landau-Lifshitz energy-momentum complex
obeys  the local conservation law 
\begin{equation}\label{LL_conserv}
L_{,\,\nu }^{\mu \nu }=0.
\end{equation}
The expressions of energy and momentum are obtained by integrating $L^{\mu\nu }$ over the 3-space
\begin{equation}\label{LL_four_mom}
P^{\mu }=\int \int \int L^{\mu 0}\,dx^{1}dx^{2}dx^{3}. 
\end{equation}
By using Gauss' theorem we have
\begin{equation}\label{LL_Gauss}
P^{\mu }=\frac{1}{16\pi }\int \int S_{,\nu }^{\mu 0i\nu }n_{i}dS=
\frac{1}{16\pi }\int \int U^{\mu 0i}n_{i}dS.
\end{equation}

The Weinberg energy-momentum complex \cite{Weinberg} is defined as
\begin{equation}\label{W}
W^{\mu \nu }=\frac{1}{16\pi }D_{,\,\lambda }^{\lambda \mu \nu },
\end{equation}
where $D^{\lambda \mu \nu }$ are the corresponding superpotentials: 
\begin{equation}\label{W_super}
D^{\lambda \mu \nu }=\frac{\partial h_{\kappa }^{\kappa }}{\partial
x_{\lambda }}\eta ^{\mu \nu }-\frac{\partial h_{\kappa }^{\kappa }}{\partial
x_{\mu }}\eta ^{\lambda \nu }-\frac{\partial h^{\kappa \lambda }}{\partial
x^{\kappa }}\eta ^{\mu \nu }+\frac{\partial h^{\kappa \mu }}{\partial
x^{\kappa }}\eta ^{\lambda \nu }+\frac{\partial h^{\lambda \nu }}{\partial
x_{\mu }}-\frac{\partial h^{\mu \nu }}{\partial x_{\lambda }},
\end{equation}
and 
\begin{equation}\label{minkowski}
h_{\mu \nu }=g_{\mu \nu }-\eta _{\mu \nu }.
\end{equation}%
The $W^{00}$ and $W^{0i}$ components correspond to the energy and momentum density, respectively. The Weinberg energy-momentum complex obeys  the local conservation law 
\begin{equation}\label{W_conserv}
W_{,\,\nu }^{\mu \nu }=0. 
\end{equation}
The integration of $W^{\mu \nu }$ over the 3-space yields the expressions
for energy and momentum:
\begin{equation}\label{W_four_mom}
P^{\mu }=\int \int \int W^{\mu 0}\,dx^{1}dx^{2}dx^{3}.
\end{equation}
By applying Gauss' theorem and evaluating the integral over the surface of a
sphere of radius $r$, the expression for the energy-momentum distribution reads:
\begin{equation}\label{W_Gauss}
P^{\mu }=\frac{1}{16\pi }\int \int D^{i0\mu }n_{i}dS.
\end{equation}

\section{Energy--Momentum Distribution of the Schwarzschild Black Hole Surrounded by
Quintessence}

Maple with GRTensor II package attached and Mathematica have been used to compute the distribution of energy and momentum. In the case of the Landau-Lifshitz and Weinberg prescriptions the calculations have to
be performed in quasi-Cartesian coordinates. Thus,  the line element given by (\ref{line_element}) is transformed in Schwarzschild Cartesian coordinates  as
\begin{equation}\label{line_element_SC}
ds^{2}=B(r)dt^{2}-(dx^{2}+dy^{2}+dz^{2})-\frac{A(r)-1}{r^{2}}(xdx+ydy+zdz)^{2},
\end{equation}
with $B(r)=1-\frac{2M}{r}-\frac{c}{r^{3w_{q}+1}}$ and $A(r)=(1-\frac{2M}{r}-\frac{c}{r^{3w_{q}+1}})^{-1}$.

In applying the Landau-Lifshitz definition we use the $U^{\mu 0i}$ quantities for the evaluation of the energy--momentum  distribution. These quantities
are involved in the expression for energy calculated by the use Gauss' theorem (\ref{LL_Gauss}).

The non-zero components of the $U^{\mu 0i}$ quantities are found to be
\begin{subequations}
\begin{equation}\label{LL_super_1}
S^{ttx}=\frac{2x}{r^{2}}\frac{(\frac{2M}{r}+\frac{c}{r^{3w_{q}+1}})}{(1-\frac{2M}{r}-\frac{c}{r^{3w_{q}+1}})}, 
\end{equation}

\begin{equation}\label{LL_super_2}
S^{tty}=\frac{2y}{r^{2}}\frac{(\frac{2M}{r}+\frac{c}{r^{3w_{q}+1}})}{(1-\frac{2M}{r}-\frac{c}{r^{3w_{q}+1}})}, 
\end{equation}

\begin{equation}\label{LL_super_3}
S^{ttz}=\frac{2z}{r^{2}}\frac{(\frac{2M}{r}+\frac{c}{r^{3w_{q}+1}})}{(1-\frac{2M}{r}-\frac{c}{r^{3w_{q}+1}})}.
\end{equation}
\end{subequations}

Substituting equations (\ref{LL_super_1})-(\ref{LL_super_3}) into (\ref{LL_Gauss}) we obtain the following expression for the energy distribution of the Schwarzschild black hole surrounded by quintessence:
\begin{equation}\label{LL_Energy}
E_{LL}=\frac{r}{2}\frac{(\frac{2M}{r}+\frac{c}{r^{3w_{q}+1}})}{(1-\frac{2M}{r%
}-\frac{c}{r^{3w_{q}+1}})}=\frac{M+\frac{c}{2r^{3w_{q}}}}{1-\frac{2M}{r}-%
\frac{c}{r^{3w_{q}+1}}},
\end{equation}
while all the momenta vanish.

In the case of the Weinberg energy-momentum complex we get the following non-vanishing components:
\begin{subequations}
\begin{equation}\label{W_super_1}
D^{xtt}=\frac{2x}{r^{2}}\frac{(\frac{2M}{r}+\frac{c}{r^{3w_{q}+1}})}{(1-%
\frac{2M}{r}-\frac{c}{r^{3w_{q}+1}})},
\end{equation}

\begin{equation}\label{W_super_2}
D^{ytt}=\frac{2y}{r^{2}}\frac{(\frac{2M}{r}+\frac{c}{r^{3w_{q}+1}})}{(1-%
\frac{2M}{r}-\frac{c}{r^{3w_{q}+1}})},
\end{equation}

\begin{equation}\label{W_super_3}
D^{ztt}=\frac{2z}{r^{2}}\frac{(\frac{2M}{r}+\frac{c}{r^{3w_{q}+1}})}{(1-%
\frac{2M}{r}-\frac{c}{r^{3w_{q}+1}})}.
\end{equation}
\end{subequations}
Inserting (\ref{W_super_1})-(\ref{W_super_3}) into (\ref{W_Gauss}) we obtain the expression for energy

\begin{equation}\label{W_Energy}
E_{W}=\frac{r}{2}\frac{(\frac{2M}{r}+\frac{c}{r^{3w_{q}+1}})}{(1-\frac{2M}{r}%
-\frac{c}{r^{3w_{q}+1}})}=
\frac{M+\frac{c}{2r^{3w_{q}}}}{1-\frac{2M}{r}-\frac{c}{r^{3w_{q}+1}}}, 
\end{equation}
while all the momenta vanish.

As it can be seen, the expression for the energy is the same in both prescriptions. The energy
distribution obtained exhibits a dependence on the mass $M$ of the black hole, the state parameter $w_{q}$ and the normalization factor $c$. 

For the special choice $w_{q}=-\frac{2}{3}$  the metric given by (\ref{special_line_element}) can
be written in Schwarzschild Cartesian coordinates  in the form (\ref{line_element_SC}) with $B(r)=1-\frac{2M}{r}-cr$ and $A(r)=(1-\frac{2M}{r}-cr)^{-1}$.
Consequently,  for the Landau-Lifshitz prescription the non-zero components of $U^{\mu 0i}$ are
\begin{subequations}
\begin{equation}\label{special_LL_super_1}
S^{ttx}=\frac{2x}{r^{2}}\frac{(\frac{2M}{r}+cr)}{(1-\frac{2M}{r}-cr)},
\end{equation}

\begin{equation}\label{special_LL_super_2}
S^{tty}=\frac{2y}{r^{2}}\frac{(\frac{2M}{r}+cr)}{(1-\frac{2M}{r}-cr)},
\end{equation}

\begin{equation}\label{special_LL_super_3}
S^{ttz}=\frac{2z}{r^{2}}\frac{(\frac{2M}{r}+cr)}{(1-\frac{2M}{r}-cr)}.
\end{equation}
\end{subequations}

Using (\ref{special_LL_super_1})-(\ref{special_LL_super_3}) and (\ref{LL_Gauss}) we obtain for the energy distribution the expression

\begin{equation}\label{special_LL_Energy}
E_{LL}=\frac{r}{2}\frac{(\frac{2M}{r}+cr)}{(1-\frac{2M}{r}-cr)}=\frac{M+\frac{cr^{2}}{2}}{1-\frac{2M}{r}-cr},
\end{equation}
while all the momenta vanish.

For the Weinberg energy-momentum complex the calculations yield
\begin{subequations}
\begin{equation}\label{special_W_super_1}
D^{xtt}=\frac{2x}{r^{2}}\frac{(\frac{2M}{r}+cr)}{(1-\frac{2M}{r}-cr)}, 
\end{equation}

\begin{equation}\label{special_W_super_2}
D^{ytt}=\frac{2y}{r^{2}}\frac{(\frac{2M}{r}+cr)}{(1-\frac{2M}{r}-cr)}, 
\end{equation}

\begin{equation}\label{special_W_super_3}
D^{ztt}=\frac{2z}{r^{2}}\frac{(\frac{2M}{r}+cr)}{(1-\frac{2M}{r}-cr)}. 
\end{equation}
\end{subequations}

From (\ref{special_W_super_1})-(\ref{special_W_super_3}) and (\ref{W_Gauss}) we get for the energy distribution inside a $2$%
-sphere of radius $r$

\begin{equation}\label{specia_W_Energy}
E_{W}=\frac{r}{2}\frac{(\frac{2M}{r}+cr)}{(1-\frac{2M}{r}-cr)}=\frac{M+\frac{cr^{2}}{2}}{1-\frac{2M}{r}-cr},
\end{equation}
while all the momenta vanish.
As expected, the expressions for the energy calculated with the Landau-Lifshitz and Weinberg complexes are the same. The energy depends on the mass of the black hole $M$ and the normalization factor $c$. 

\section{Discussion}

In the present work the energy-momentum distribution for a four-dimensional\break  Schwarzschild black hole surrounded by quintessence is studied by use of the Landau-Lifshitz and Weinberg energy-momentum complexes. In both prescriptions the expression for the energy obtained is the same depending on the mass $M$ of the black hole, the state parameter $w_{q}$ and a positive normalization factor $c$, while all the momenta vanish. The energy distribution is also given for the special choice $w_{q}=-\frac{2}{3}$. In the absence of quintessence, the energy in Schwarzschild Cartesian coordinates becomes $E_{LL}=E_{W}=M(1-\frac{2M}{r})^{-1}$ in agreement with the result obtained in \cite{Virbhadra_2}.

Two limiting cases are of interest for the special choice $w_{q}=-\frac{2}{3}$ and these are presented in the following Table:
\[
\begin{tabular}{lc}
Limit & Energy $E_{LL}=E_{W}$ \\ 
$r\rightarrow 0$ & $0$ \\ 
$r\rightarrow \infty $ & $-\infty $
\end{tabular}
\]

Further, we consider the energy calculated in \cite{Saleh} by applying the Einstein and M\o ller energy-momentum complexes for a Reissner-Nordstr\"om black hole surrounded by quintessence. This energy depends on the charge $Q$ and the mass $M$ of the black hole, the state parameter $w_q$ and the normalization factor $c$. For the case of a Schwarzschild black hole ($Q = 0$), the comparison with our results is presented in the following table:
\[
\begin{tabular}{lc}
\textbf{Energy-Momentum Complex} & \textbf{Energy} \\ 
&\\
Landau-Lifshitz & $E_{LL}=\displaystyle{\frac{M+\frac{c}{2r^{3w_{q}}}}{1-\frac{2M}{r}-%
\frac{c}{r^{3w_{q}+1}}}}$ \\ 
&\\
Weinberg & $E_{W}=\displaystyle{\frac{M+\frac{c}{2r^{3w_{q}}}}{1-\frac{2M}{r}-%
\frac{c}{r^{3w_{q}+1}}}}$  \\ 
&\\
Einstein &  $E_{E}=M+\displaystyle{\frac{c}{2}\frac{1}{r^{3w_{q}}}}$ \\
&\\
M\o ller & $E_{M}=M+\displaystyle{\frac{c}{2}\frac{3w_{q}+1}{r^{3w_{q}}}}$
\end{tabular}
\]

As it is evident, the presence of quintessence around the black hole influences the energy of the gravitational field of the latter. 
Certainly, the results obtained in the present work do not settle the issue of the energy-momentum localization in General Relativity. However, they do contribute to the ongoing debate on the problem. It is expected that further investigations concerning other dark energy candidates around black holes and/or the application of other energy-momentum complexes will shed more light on the issue.

\end{document}